\documentclass[pra,aps,twocolumn,showpacs]{revtex4}
\usepackage{amsfonts,amssymb,graphics,graphicx,epsfig,color,times,bbm}
\usepackage[latin1]{inputenc}
\usepackage[intlimits,centertags]{amsmath}

\graphicspath{{Bilder/}}

\begin{document}
 
\title{Analytic approximations to the phase diagram of the Jaynes-Cummings-Hubbard model
with application to ion chains}
\author{Alexander Mering}
\affiliation{Fachbereich Physik and research center OPTIMAS, Technische Universit\"at Kaiserslautern, D-67663 Kaiserslautern, Germany}
\email{amering@physik.uni-kl.de}
\author{Michael Fleischhauer}
\affiliation{Fachbereich Physik and research center OPTIMAS, Technische Universit\"at Kaiserslautern, D-67663 Kaiserslautern, Germany}
\author{Peter A. Ivanov}
\affiliation{Institut f\"ur Quanteninformationsverarbeitung, Universit\"at Ulm, Albert-Einstein-Allee 11, 89081 Ulm, Germany}
\author{Kilian Singer}
\affiliation{Institut f\"ur Quanteninformationsverarbeitung, Universit\"at Ulm, Albert-Einstein-Allee 11, 89081 Ulm, Germany}
\begin{abstract}
We discuss analytic approximations to the ground state phase diagram of the 
homogeneous Jaynes-Cummings-Hubbard (JCH) Hamiltonian
with general short-range hopping. The JCH model describes e.g. radial phonon excitations of a linear chain of ions coupled to
an external laser field tuned to the red motional sideband with Coulomb mediated hopping or an array of high-$Q$ coupled
cavities containing a two-level atom and photons. Specifically we consider 
the cases of a linear array of coupled cavities and a linear ion chain. We 
derive approximate analytic expressions for the boundaries between Mott-insulating and superfluid
phases and give explicit expressions for the critical value of the hopping amplitude within the different 
approximation schemes. In the case of an array of cavities, which is represented by the 
standard JCH model we compare both approximations to numerical data from density-matrix renormalization group (DMRG) calculations.
\end{abstract}


\date{\today}

\maketitle

%


In recent years there is a growing interest in quantum optics systems as experimental
testing ground of fundamental models for quantum many body physics and quantum
simulation. The most prominent example are certainly ultra-cold atoms 
in optical lattices \cite{Bloch2008, Greiner2002}, which are almost ideal representations of various types 
of Hubbard models \cite{Essler2008,Jaksch1998,Albus2003,Schneider2008,Joerdens2008}. However due to their finite mass, atomic systems represent,
with few exceptions, only models with explicit particle number conservation. 
On the other hand different quantum optical systems employing photons or
quasi-particles such as phonons have been suggested recently, which are
not limited by this constraint. For example an array of coupled high-$Q$ micro-cavities 
containing a two-level atom and a photon is described by the Jaynes-Cummings-Hubbard model (JCHM) 
\cite{Hartmann2006,Hartmann2008,Hartmann2008a,Angelakis2007}.
It is a combination of two well-known systems, the Jaynes-Cummings model \cite{Jaynes1963,Shore1993}
describing the
coupling of a single two-level system to a bosonic mode and the hard-core Bose-Hubbard model
\cite{Fisher1989}
which describes the interaction and tunneling of bosons on a lattice. Recently
we have shown that a modification of the JCHM can also be implemented 
in a linear ion trap, which has the advantage of an easier experimental
realization since the required parameter regime is already realizable with current technology \cite{Ivanov2009}. A large variety of analytic and numeric methods was applied to the JCHM and related models, providing profound results for the phase diagram and other ground state quantities \cite{Koch2009,Pippan2009,Makin2008,Aichhorn2008,Rossini2007,Lei2008,Schmidt2009,Greentree2006}. In the present paper we show that in the strong interaction limit and near commensurate filling
simple approximate analytic solutions of the JCHM can be found if 
there is translational invariance, i.e. for
an infinite homogeneous system or periodic boundary conditions realizable e.g. with ions in a race-track Paul trap design.
These solutions provide a good analytic approximation to the full ground state phase diagram.

This paper is structured as followed. In section \ref{sec:intro} we will briefly
summarize the main properties of the Jaynes-Cummings model, together with other important quantities 
needed later on. In section \ref{sec:Approximations} we introduce two different approximation schemes, both giving analytic results for the critical hopping amplitude for the Mott insulator (MI) to superfluid (SF) transition. In section \ref{sec:Application}
we apply both approximation to the simple cubic nearest-neighbor JCH model describing an array of coupled
cavities and to the special case of an linear ion chain.

\section{the JCH model}\label{sec:intro}
In this section, we will shortly review the main features of the JCH model defined by the Hamiltonian
\begin{multline}
  \hat H = \omega \sum_{ j} \hat a_{ j}^\dagger \hat a_{ j} + \Delta \sum_{ j} 
\hat\sigma^\dagger_{ j} \hat\sigma^-_{ j} + g\sum_{ j} \left(\hat\sigma ^\dagger_{ j} \hat a_{ j} + \hat a_{ j}^\dagger \hat\sigma_{ j}^-\right)\\
   +\sum_{d} t_{ d} \sum_{ j} \left(\hat a^\dagger_{ j} \hat a_{ j+ d} + \hat a^\dagger_{ j+ d} \hat 
a_{ j}\right)\label{eq:JCHreal}
\end{multline}
and discuss the main quantities needed in order to calculate its phase diagram. The system (\ref{eq:JCHreal}) comprises
bosonic and spin degrees of freedom, the specific interpretation of which depends on the actual physical system. 
Depending on the implementation, $\hat a_j^\dagger$ and $\hat a_j$ describe the creation and annihilation of a photon (phonon) at the $j$th cavity (ion),
$\hat \sigma_j^\pm$ are the spin flip operators between the internal states of the atom (ion) and $\Delta$ is the transition energy of the atom (the detuning
of the external laser field from the red motional sideband). $g$ describes the cavity-mediated atom-photon coupling (the phonon-ion coupling in the Lamb Dicke limit) and $\omega$ is the cavity resonance (the local oscillation) frequency. Between separated cavities (ions), there is a photon (phonon) transfer described in (\ref{eq:JCHreal}) by the distance dependent hopping amplitude $t_d$. 
\\

In the limit of vanishing hopping $t_d\equiv0$, the resulting Jaynes-Cummings model can easily be diagonalized. In this case, all sites $j$ decouple and become independent. Since the total number of excitations $\hat N_j = \hat a_j^\dagger \hat a_j + \hat \sigma^+_j \hat \sigma^-_j$ on every site $j$ is a constant of motion, the local JC Hamiltonian block diagonalizes. Within each two-dimensional excitation subspace, the eigenstates can easily be found. Adapting the notation from \cite{Makin2008}, these are given by
\begin{align}
 \left| \pm , n\right \rangle &= \frac{\left[ \chi_n\mp(\omega-\Delta)\right]   \left| \uparrow , 
n-1\right\rangle \pm 2 g\sqrt n   \left| \downarrow , n\right\rangle}{\sqrt{2}\sqrt{\chi_n^2\mp (\omega-\Delta)\chi_n}}\label{eq:JCGroundState}\\
 &:= \alpha_n^\pm  \left| \uparrow , n-1\right\rangle   \pm \beta_n^\pm \left| \downarrow , n\right\rangle
\end{align}
with $\chi_n=\sqrt{(\Delta-\omega)^2+4 n g^2}$ and $n>0$, and the eigenenergies are
\begin{equation}
 E_n^\pm = n \omega +\frac{\Delta-\omega}2 \pm \frac12\chi_n. \label{eq:JCEnergy}
\end{equation}
For $n=0$, the ground state is non-degenerate and given by  $\left|-,0\right\rangle =\left|\downarrow ,0\right \rangle$ with $E_0=0$.  Here, the state $ \left| \uparrow , n-1\right\rangle$ describes an atomic excitation together with $n-1$ bosonic excitations; $\left| \downarrow , n\right\rangle$ is  the state with the atom in the ground state and $n$ bosonic excitations. In the strong interaction limit $g\gg |\Delta-\omega|$,  the energy gap $\Delta E_n = E_n^+ - E_n^- = \chi_n\sim 2 g\sqrt n$ is large compared to any other energy scale in the system and thus the excited states $ \left|+ , n\right \rangle$ do not contribute to the ground state.\\

For the following discussion it will be useful to consider the action of a single bosonic creation or annihilation 
operator on a given JC eigenstate  $\left| \pm , n\right \rangle$. Defining 
\begin{align}
 A^\pm_n&=\begin{cases} \sqrt n\ \alpha^\pm_n\beta^-_{n+1} \pm\sqrt{n+1}\ \beta_n^\pm\alpha^-_{n+1}
& n>0\\\hspace{2cm} \alpha^-_1 &n=0\end{cases}\label{eq:factorsA}\\
 B^\pm_n&=\begin{cases}\sqrt n \ \alpha^\pm_n\beta^+_{n+1} \mp\sqrt{n+1}\ \beta_n^\pm\alpha^+_{n+1}& n>0\\\hspace{2cm} 
-\alpha^+_1 &n=0\end{cases}\\
 C^\pm_n&=\begin{cases} \sqrt {n-1}\ \alpha^\pm_n\beta^-_{n-1} \pm\sqrt{n}\ \beta_n^\pm\alpha^-_{n-1}
& n>1\\ \hspace{2cm} 0&n\le1\end{cases}\\
 D^\pm_n&=\begin{cases} \sqrt {n-1}\  \alpha^\pm_n\beta^+   _{n-1} \mp\sqrt{n}\ \beta_n^\pm\alpha^+_{n-1}& n>1
\\\hspace{2cm} \pm \beta^\pm_1 \delta_{n,1}&n\le1\end{cases},\label{eq:factorsD}
\end{align}
the action of $\hat a^\dagger $ and $\hat a$ on the state $\left| \pm , n\right \rangle$ can be seen to be
\begin{align}
\hat a^\dagger  \left| \pm , n\right \rangle 	&=  A^\pm_n\left| + , n+1\right\rangle  +  B^\pm_n  \left| - , n+1\right  \rangle\\
 \hat a \left| \pm , n\right \rangle			&=  C^\pm_n \left| + , n-1\right\rangle +  D^\pm_n\left| - ,
 n-1\right  \rangle   ,
\end{align}
i. e.  $\hat a^ \dagger$ and $\hat a$ connect the manifold of states $\left| \pm , n\right  \rangle$ to the manifolds $\left| \pm , n+1\right  \rangle$  and $\left| \pm , n-1\right  \rangle$ respectively as expected.\\

In order to calculate the phase boundaries of the Mott insulating lobes for the JCH model, we will follow the usual route. Since the total number of excitations in the system
\begin{equation}
 \hat N=\sum_{   j} \left( \hat a_{   j}^\dagger\hat a_{   j}+ \hat\sigma^+_{   j} \hat\sigma^-_{   j}\right)
\end{equation}
commutes with the full Hamiltonian (\ref{eq:JCHreal}), it is enough to treat the system for a fixed number of excitations. The boundary of the $n$th Mott lobe can then be determined by calculating the total energy $E(N)$ for $N=nL-1$, $N=nL$ and $N=nL+1$ excitations  in a system with $L$ sites. The chemical potential then reads
\begin{equation}
\mu^\pm_n = \pm\Bigl[ E({n L\pm1})- E({n L})\Bigr] ,\label{eq:chemicalPotential} 
\end{equation}
where the plus sign belongs to the upper boundary of the Mott lobe and the minus sign to the lower one. For $t_d\equiv0$ ,
$\mu^ \pm_n$ can be calculated 
straight forwardly. Starting with the energy for $N=nL$ excitations with $n$ being an integer, i.e. for a commensurate number of excitations, it can be seen that due to the nonlinear dependence of the single-site energy $E^-_n$ on $n$, the excitations will distribute equally over the whole lattice. The ground state is therefore given by $\vec n=\{n,n,\dots,n\}$. Now, when adding (removing)  a single excitation from the whole system, the ground state is given by $\{n\pm1,n,\dots,n\}$, where we have ignored the degeneracy of the state since we are only interested in the energy and the system is homogeneous. With this, the energies at $t_d=0$ can be written as
\begin{align}
 E(nL-1)  &= (L-1) E^-_n +  E^-_{n-1} \\
 E(nL)      &=L E^-_n \\
 E(nL+1) &=(L-1) E^-_n +  E^-_{n+1},
\end{align}
and the chemical potentials evaluate to
\begin{align} 
\mu^+_n&=E^-_{n+1}-E^-_n \\
 &= \omega - \frac {\chi_{n+1}} 2 +(1-\delta_{n0}) \frac {\chi_{n}} 2+\delta_{n,0} \frac{\Delta-\omega}2,\\
 \intertext{for any $n$ and}
 \mu^-_n&= E^-_n-E^-_{n-1} \\
 &= \omega - \frac {\chi_{n}} 2 +(1-\delta_{n1}) \frac {\chi_{n-1}} 2+\delta_{n,1} \frac{\Delta-\omega}2,
\end{align}
for $n>0$.  
Thus for a commensurate number of excitations the system displays particle-hole gaps. 
Since $\mu^-_{n+1}=\mu^+_n$ the chemical potential for  non-commensurate total
number of excitations between $N=nL$ and $N=(n+1)L$ is the same, corresponding to a critical
point. For non-vanishing tunneling the critical points extend to critical regions. \\

The simplest numerical method to obtain a qualitative phase diagram is the so called mean field theory. As described for instance in \cite{Greentree2006,Hartmann2008a,Lei2008,Makin2008}, mean field theory can be implemented by introducing an order parameter $\Psi$, which in our case is chosen to be homogeneous and real valued. Decoupling the hopping term by using
\begin{equation}
 \hat a_j^\dagger \hat a_l \mapsto \Psi  \left(\hat a_j^\dagger + \hat a_l\right) - \left|\Psi\right|^2,
\end{equation}
the whole JCH Hamiltonian (\ref{eq:JCHreal}) in the grand-canonical ensemble uncouples in real space with a local Hamiltonian being
\begin{multline}
 \hat H^{\rm MF} = \left(\omega-\mu\right)  \hat a^\dagger \hat a+ \left(\Delta-\mu\right) \hat \sigma^+\hat\sigma^- + g\left(\hat a^\dagger \hat\sigma^- + \hat a\hat \sigma^+\right)\\
 - 2 \widetilde J \Psi  \left(\hat a^\dagger + \hat a\right) + 2\widetilde J \left|\Psi\right|^2.\label{eq:JCHMeanField}
\end{multline}
At this point, we omitted the spatial index because the problem is purely local. The modified hopping amplitude $\widetilde J=-\sum_d t_d$ gives the effective coupling within the mean field scheme. The phase diagram is now found by diagonalizing the mean field Hamiltonian (\ref{eq:JCHMeanField}) either exactly by means of perturbation theory or numerically, setting an upper bound for the maximal number of bosonic excitations in the system. The ground state energy is then given by $\underset{\Psi}{\min}\  E[\Psi]$ and the MI is distinguished from the SF by the value of $\Psi$ for the minimal energy. For $\Psi\equiv0$, the system is in a MI state, for $\Psi>0$, the ground state is superfluid. This sets the point of the MI to SF transition. It should be mentioned at this point, that this method gives inadequate results in one dimension ($D=1$)  but is exact for $D\to\infty$. Additionally, the effective hopping $\widetilde J$ must be larger than zero to yield useful results.

\section{Approximative determination of the phase boundaries}\label{sec:Approximations}

\subsection{Effective strong-coupling model}\label{sec:effective}

From the discussion above, it can be seen that the phase boundaries are defined by the closure of the particle-hole gap. In the present subsection, we will derive effective Hamiltonians in the strong-coupling limit for the calculation of the upper and lower chemical potential of the $n$th Mott lobe, allowing to calculate the particle-hole gap in first order of the hopping amplitudes $t_d$.  To do so, we employ degenerate perturbation theory using Kato's expansion as summarized in \cite{Klein1974} in first order with $H_{\rm eff} = \mathcal P V\mathcal P$. This procedure is equivalent to the polariton mapping considered in \cite{Angelakis2007,Schmidt2009}.  First, we note that according to eq. (\ref{eq:JCEnergy}) the state $\left| + , n\right \rangle $ is separated by a large energy gap from the ground state $\left|- , n\right \rangle $. Thus,  $\left| + , n\right \rangle $ can be completely neglected in the following as already mentioned in  \cite{Angelakis2007,Koch2009}. \\

When looking for the energy of the ground state with $N=nL$, from perturbation theory, no 1st order contributions are present. So, the Hilbert space per site is one dimensional, consisting of the single state $\left|- , n\right \rangle$. Thus, up to first order, the energy is given by  $E(nL)  =L E^-_n$. When adding an excitation, the local Hilbert space increases; now, (locally), the two states $\left|- , n\right \rangle$ and $\left|- , n+1\right \rangle$ need to be taken into account. So, in this limit, the system for an additional particle can be understood as a system consisting of effective spin $\frac 12$ particles. We will identify the states $\left| \Uparrow\right\rangle$ with the state $\left|- , n+1\right \rangle$ and $\left| \Downarrow\right\rangle$ with $\left|- , n\right \rangle$. In order to derive the effective spin $\frac12$ model, one has to look on the action of the hopping operator $\hat a_{j+1}^\dagger \hat a_j$  on the states in the Hilbert space. Using equations (\ref{eq:factorsA}) to (\ref{eq:factorsD}) and neglecting the contributions from the states  $\left|+ , n\right \rangle$ and $\left|+ , n+1\right \rangle$, the hopping operator $\hat a^\dagger_{j+1} \hat a_j$ acts as
\begin{equation}
 \hat a^\dagger_{j+1} \hat a_j\left|\Downarrow\right \rangle_{j+1}\left|\Uparrow\right \rangle_j = B^-_n 
D^-_{n+1}\left|\Uparrow\right \rangle_{j+1}\left|\Downarrow\right \rangle_j
\end{equation}
within the considered subspace. Therefore, by introducing spin operators ${\widetilde\sigma}^\pm_j$ the hopping term is equivalent to a nearest neighbor spin-spin interaction with
\begin{equation}
 \hat a^\dagger_{j+1} \hat a_j = B^-_n D^-_{n+1} {\widetilde\sigma}^+_{j+1} {\widetilde\sigma}^-_{j} .
\end{equation}
Together with the energy of the system, one can thus write an effective Hamiltonian 
describing the upper boundary of the $n$th Mott lobe
\begin{multline}
{ \widetilde  H} = E^-_{n} \sum_j {\widetilde\sigma}^-_{j} {\widetilde\sigma}^+_{j}  +E^-_{n+1} \sum_j {\widetilde\sigma}^+_{j} {\widetilde\sigma}^-_{j} \\
 +B^-_n D^-_{n+1}  \sum_d t_d \sum_j \left( {\widetilde\sigma}^+_{j+d} {\widetilde\sigma}^-_{j}+ 
{\widetilde\sigma}^+_{j} {\widetilde\sigma}^-_{j+d}\right).
\end{multline}
This Hamiltonian is equivalent to 
\begin{multline}
{\widetilde  H} = (L-1) E^-_{n}  +E^-_{n+1}  \\
 +B^-_n D^-_{n+1}  \sum_d t_d \sum_j \left( {\widetilde\sigma}^+_{j+d} {\widetilde\sigma}^-_{j}+ 
{\widetilde\sigma}^+_{j} {\widetilde\sigma}^-_{j+d}\right)
\end{multline}
since we are at fixed magnetization with only one spin pointing upwards. This Hamiltonian can be further simplified, by using a Jordan-Wigner transformation mapping the spin operators ${\widetilde\sigma}^-_{j}$ onto fermionic operators $\hat c_j$ and subsequently performing a Fourier transformation
\begin{equation}
  \hat c_{   j} = \frac{1}{\sqrt{L}}\sum_{   k} e^{-2\pi i\frac{   k   j}{L}} \hat c_{   k}.
 \end{equation}
Then, the ground state wave function factorizes, since the Hamiltonian decouples in momentum space
\begin{multline}
   {\widetilde  H} = (L-1) E^-_{n}  +E^-_{n+1} \\
   + 2 B^-_n D^-_{n+1}  \sum_d t_d \sum_k \cos(2\pi\frac{   k    d}L) \hat c_k^\dagger \hat c_k.
 \end{multline}
This model is equivalent to free fermionic particles with hopping amplitudes given by $t_d$. In momentum space, a
single fermion will occupy the mode with lowest energy. Thus the total energy of the single particle, and therefore the total energy of an additional excitation on top of the $n$th Mott insulator in the JCH model is given by
\begin{multline}
  E(nL+1) =   (L-1)  E^-_{n}  +E^-_{n+1}  +    F_n(k^\prime), \label{eq:effectiveEnergyPlus}
\end{multline}
where 
\begin{equation}
 F_n(k)=   2\  B^-_n D^-_{n+1} \sum_d t_d \cos(2\pi\frac{   k  d}L)
\end{equation}
 and  the momentum mode  $k^\prime$ is chosen such that $F_n(k)\bigr|_{k^\prime}$ is minimal. It should be mentioned that 
the product $B_n^- D_{n+1}^-$ is positive for any $(\Delta,\omega,n)$, so the momentum mode is purely determined by the minimum of $\sum_d t_d \cos(2\pi\frac{   k  d}L)$.\\

To calculate the energy for a hole in the $n$th Mott insulator, we follow exactly the same route. Now, the state $\left| \Downarrow\right\rangle$ is associated with $\left|- , n-1\right \rangle$ and  $\left| \Uparrow\right\rangle$ with $\left|- , n\right \rangle$. The hopping operators act as
\begin{equation}
 \hat a^\dagger_{j+1} \hat a_j\left|\Downarrow\right \rangle_{j+1}\left|\Uparrow\right \rangle_j = B^-_{n-1} 
D^-_{n}\left|\Uparrow\right \rangle_{j+1}\left|\Downarrow\right \rangle_j
\end{equation}
and the effective Hamiltonian is given by
\begin{multline}
{ \widetilde  H} = E^-_{n-1} \sum_j {\widetilde\sigma}^-_{j} {\widetilde\sigma}^+_{j}  +E^-_{n} \sum_j {\widetilde\sigma}^+_{j} {\widetilde\sigma}^-_{j} \\
 + B^-_{n-1} D^-_{n}\sum_d t_d \sum_j \left({\widetilde\sigma}^+_{j+d} {\widetilde\sigma}^-_{j}+
{\widetilde\sigma}^+_{j} {\widetilde\sigma}^-_{j+d}\right).
\end{multline}
Here the magnetization consist of one spin pointing downward. Again, after making use of a Jordan-Wigner transformation and subsequently a Fourier transformation, the energy of a
single hole is given by
 \begin{multline}
  E(nL-1) =    (L-1) E^-_{n}  +E^-_{n-1}+F_{n-1}(k^{\prime\prime})  , \label{eq:effectiveEnergyMinus}
 \end{multline}
 where the same condition holds for $k^{\prime\prime}$. Now, putting the calculated energies 
(\ref{eq:effectiveEnergyPlus}) and (\ref{eq:effectiveEnergyMinus}) together, the chemical potentials and therefore the boundaries of the $n$th Mott insulating lobe can easily be derived. They are given by
 \begin{align}
  \mu^+_n&= E^-_{n+1}  - E^-_{n}  +   2 B^-_n D^-_{n+1}  \sum_d t_d \cos(2\pi\frac{   k^\prime    d}L)\label{eq:chemicalPotentialfirstPlus}\\
  \mu^-_n &=E^-_{n}   - E^-_{n-1}  -   2  B^-_{n-1} D^-_{n} \sum_d t_d \cos(2\pi\frac{   k^{\prime\prime}    d}L)\label{eq:chemicalPotentialfirstMinus} .
 \end{align}
where $k^\prime$ ($k^{\prime\prime}$) is chosen such that $\mu_n^+(k^\prime)$ ($\mu_n^-(k^{\prime\prime})$) is minimal (maximal). This result generalizes the findings from \cite{Schmidt2009,Koch2009} to arbitrary hoppings $t_d$.

\subsection{Fermion approximation}\label{sec:commapprox}
In this subsection, we will apply an even simpler, but not that obvious approximation. When looking at the JCH Hamiltonian (\ref{eq:JCHreal}), it can be seen that all terms are quadratic. These kinds of models are in general suited for an exact solution by means of a Fourier transform. The problem at this point is however, that the commutation relations of
spin operator $\hat\sigma_j^\pm$ are not as simple as that of bosons or fermions.
 This limits the applicability of a Fourier transform, since the operators in momentum space will not obey the same 
commutation relation as in real space \footnote{The authors would like to thank E. Irish and M. Hartmann for pointing out a
corresponding error in a previous version of this manuscript.}. The usual step of a prior Jordan-Wigner transformation, transforming the spin operators to proper fermionic operators, is not applicable in this case, since the interaction part is linear in the spin operators, so the Jordan-Wigner factors do not cancel out. Thus both transformations cannot be carried out exactly without increasing the descriptional complexity of the problem. Nevertheless the Hamiltonian can be
diagonalized by a Fourier-transform in an approximate way.
\\

As said above, all modes decouple at $t_d=0$. For this reason, the spin-operators are in this limit  equivalent to fermionic operators. If we assume that this replacement also holds for small values of $t_d$, the JCH model (\ref{eq:JCHreal}) can be rewritten in a fermionic approximation  
\begin{multline}
  \hat H = \omega \sum_{ j} \hat a_{ j}^\dagger \hat a_{ j} + \Delta \sum_{ j} 
\hat c^\dagger_{ j} \hat c_{ j} + g\sum_{ j} \left(\hat c ^\dagger_{ j} \hat a_{ j} + \hat a_{ j}^\dagger \hat c_{ j}\right)\\
   +\sum_{d} t_{ d} \sum_{ j} \left(\hat a^\dagger_{ j} \hat a_{ j+ d} + \hat a^\dagger_{ j+ d} \hat 
a_{ j}\right).
\end{multline}
Here the spin operators $\hat\sigma^+$ ($\hat \sigma^-$) are replaced by fermionic operators $\hat c^\dagger$ ($\hat c$). Within this approximation, a Fourier transform of both, the bosonic and fermionic degrees of freedom can be easily accomplished via
\begin{align}
  \hat a_{   j} &= \frac{1}{\sqrt{L}}\sum_{   k} e^{-2\pi i\frac{   k    j}{L}} \hat a_{   
k}\label{eq:FourierBoson}\\
 \hat c_{   j} &= \frac{1}{\sqrt{L}}\sum_{   k} e^{-2\pi i\frac{   k   j}{L}} \hat c_{   
k}\label{eq:FourierFermion}.
\end{align}
Here $\hat a_{   k}$ and $\hat c _{   k}$ are operators in momentum space. Doing so,  the JCH Hamiltonian transforms to that of uncoupled Jaynes-Cummings systems
\begin{gather}
  \hat H= \sum_{   k}  \omega_{   k}  \hat a_{   k} ^\dagger \hat a_{   k}  + \Delta \sum_{   k} \hat 
 c^\dagger_{   k}  \hat c_{   k}  + g\sum_{   k}  \left( \hat c ^\dagger_{   k}  \hat a_{   k}  + \hat a_{   k}^\dagger 
\hat c_{   k}\right),\label{eq:JCHMomentum}
\end{gather}
with
\begin{gather}
\omega_k = \omega + 2 \sum_{   d} t_{   d} \cos(2\pi\frac{   k    d}L). \label{eq:PhononEnergies}
\end{gather}
The ground state in any mode is given by the Jaynes-Cummings ground state (\ref{eq:JCGroundState}) with frequency $\omega_k$. The energy of mode $   k$ with $n$ excitations is
\begin{equation}
 E^{n}_{   k} = (1-\delta_{n0})\left[ n\, \omega_{   k} + \frac{\Delta-\omega_{   k}}{2}-\frac 12\sqrt{ (\Delta-\omega_{   
k})^2+4 n g^2}\right]. \label{eq:JCEnergyMomentum}
\end{equation}

Since the total number of excitations in the system
\begin{equation}
 \hat N=\sum_{   j} \left( \hat a_{   j}^\dagger\hat a_{   j}+ 
\hat \sigma^\dagger_{   j} \hat \sigma^-_{   j}\right)\mapsto\sum_{   
k} \left( \hat a_{   k}^\dagger\hat a_{   k}+  \hat c^\dagger_{   k} \hat c_k\right)
\end{equation}
commutes with the Hamiltonian (\ref{eq:JCHMomentum}), a common basis can be chosen.  Thus the full solution of (\ref{eq:JCHMomentum}) for a fixed total number of excitations $N=nL$ is given by the distribution 
$ \vec {n}=\{n_{   k_1},n_{   k_2},\dots\}$ of $N$ excitations on $L$ momentum modes with minimal energy 
$E_N[  \vec n]=\sum_{   k} E^{n_{   k}}_{   k}$ together with the constraint $\sum_{   k} n_{   k}\equiv N$. 
Note that the number of momentum modes $L$ is equal to the number of sites. \\

When constructing the phase diagram, the energy of $N=nL-1$, $N=nL$ and $N=nL+1$ excitations needs to be calculated.
In the limit of vanishing hopping ($t=0$) and for commensurate filling, i.e.
$N=nL$, the distribution of occupation numbers
which has the lowest energy is again $\vec n =\{n,n,\dots,n\}$. This corresponds
to a MI state with an integer number of excitations on every lattice sites. The
phase is gapped with a particle-hole gap as described in section \ref{sec:intro}. When $t$ is increased the ground state
remains the same, but the gap closes and a quantum phase transition occurs from the
MI to the SF phase at some critical value of $t$. The only remaining thing in order to calculate the chemical potentials is to find the momentum mode where the addition (removal) of an excitation gives the maximum (minimum) reduction (increase)  in the total energy. This yields
\begin{align}
  \mu_n^+ &=  E_{k^\prime}^{n+1}-E_{k^\prime}^{n}\label{eq:ChemicalPlusCommutator}\\
  \mu_n^- &=  E_{k}^{n}-E_{k}^{n-1},\label{eq:ChemicalMinusCommutator}
\end{align}
where $k^\prime$ ($k$) is chosen such that $\mu_n^+(k^\prime)$ ($\mu_n^-(k)$) is minimal (maximal). The actual values of $k$ and $k^\prime$ depend mainly on the sign of the hopping amplitudes $t_d$.

\section{Application to specific realizations of the JCH model}\label{sec:Application}

After having introduced the two approaches used in this paper, we will apply them to the case of the simple 
JCHM with positive effective mass and nearest neighbor hopping and to a modified model describing the physics of a linear ion chain. The case of the simple JCHM essentially serves as a testing ground for our approximation schemes including
a comparison of the analytic results to numerical data from DMRG and mean field calculations.  Later on, the generalized JCHM will be treated by both approximations giving analytic results for the phase diagram in a wide range of parameters.

\subsection{JCHM with positive effective mass and nearest-neighbor hopping}

Without loss of generality,  we will specialize here on the case discussed in \cite{Rossini2007}. The Hamiltonian of the JCHM in this case is given by
\begin{multline}
 \hat H = \omega \sum_{ j} \hat a_{ j}^\dagger \hat a_{ j} + \Delta \sum_{ j} \hat\sigma^\dagger_{ j} \hat\sigma^-_{ j} + 
g\sum_{ j} \left(\hat\sigma ^\dagger_{ j} \hat a_{ j} + \hat a_{ j}^\dagger \hat\sigma_{ j}^-\right)\\
 -t \sum_{ j} \left(\hat a^\dagger_{ j} \hat a_{ j+ 1} + \hat a^\dagger_{ j+ 1} \hat 
a_{ j}\right).\label{eq:JCH}
\end{multline}
with $\omega=\Delta$. Comparing with the Hamiltonian eq. (\ref{eq:JCHreal}), one notes that 
the hopping amplitudes satisfy $t_d = - t \delta_{d1}$.\\

For the calculation of the chemical potentials, we first have to determine the momentum modes $k^\prime$ and $k^{\prime\prime}$ which contribute to the energy. For $\omega=\Delta$, the coefficients in (\ref{eq:JCGroundState}) are $\alpha_n^\pm = \frac 1{\sqrt 2} = \beta_n^\pm$ and therefore
\begin{equation}
 B_n^- = \left \{ \begin{matrix}\frac{\sqrt n+\sqrt{n+1}}{2} &&\hspace{1cm} n>0\\ -\frac{1}{\sqrt 2} &&\hspace{1cm} n=0 
\end{matrix}\right\} = D_{n+1}^-.
\end{equation}
With this, the function $F_n(k)$ is given by 
\begin{equation}
 F_n(k) = - t \frac{\left(\sqrt n+\sqrt{n+1}\right)^2}{2-\delta_{n,0}} \cos(2\pi \frac kL).
\end{equation}
The both chemical potentials have its minimum (maximum) at $k=0$. Putting everything together, the phase boundaries of the $n$th Mott lobe, 
calculated using the effective strong-coupling model read
\begin{align}
 \mu^+_n &= \omega-\frac 12 \chi_{n+1} + \frac{1-\delta_{n0}}2 \chi_n -t\frac{\left(\sqrt 
n+\sqrt{n+1}\right)^2}{2-\delta_{n0}}, \\
\intertext{for any $n$ and}
 \mu^-_n &= \omega-\frac 12 \chi_{n} + \frac{1-\delta_{n1}}2 \chi_{n-1} +t\frac{\left(\sqrt 
n+\sqrt{n-1}\right)^2}{2-\delta_{n1}},
\end{align}
for $n>0$.  This allows for the determination of the critical hopping amplitude $t_{\rm crit}$ where $ \mu^+_n= \mu^-_n$, which is given by 
\begin{equation}
 t_{\rm crit}/g = 2\frac{2 \sqrt n - \sqrt{n+1}-\sqrt{n-1}}{(\sqrt{n}+\sqrt{n+1})^2+(\sqrt{n+\delta_{n1}}+\sqrt{n-1})^2}.
 \end{equation}

\begin{figure}
\epsfig{file=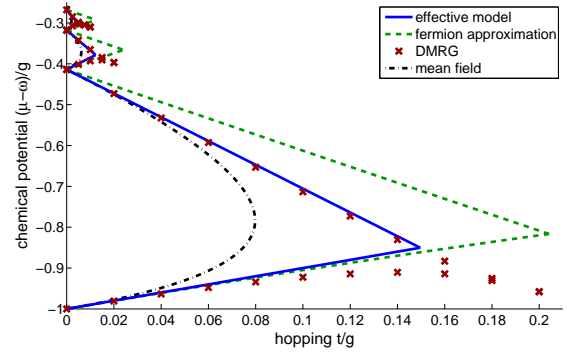,width=1\columnwidth}
\caption{(Color Online) Comparison of ground state phase diagram of the 1D JCHM (\ref{eq:JCH}) obtained by DMRG
(red crosses, from Rossini and Fazio PRL \textbf{99} 186401 (2008)) as well as mean field
results (dot-dashed line) with the prediction from our approaches 
(solid line: strong-coupling effective Hamiltonian; dashed line: fermion approximation)
for $\Delta=\omega=1$ and $g=1$. 
Taking into account the simplicity of both
approaches, the agreement with the DMRG data is rather good while the mean field predictions are
rather poor as expected for 1D systems. The critical hopping amplitudes estimated from the DMRG data 
agree surprisingly well with those predicted within the fermion approximation, although the shape of the Mott lobe is different. }\label{fig:DMRG}
\end{figure}

Secondly we apply the second approximation to this model. With the given system parameters, the momentum dependent phonon energies (\ref{eq:PhononEnergies}) are given by
\begin{gather}
\omega_k = \omega - 2  t \cos(2\pi\frac{   k  }L)
\end{gather}
and the energy in the $k$th momentum mode for a given filling $n$ reads (see eq. (\ref{eq:JCEnergyMomentum}))
\begin{multline}
 E^{n}_{   k} = (1-\delta_{n0})\Biggl[n\omega-2 n\,  t \cos(2\pi\frac{   k  }L) + t \cos(2\pi\frac{   k  }L)\\
    -\sqrt{ t^2 \cos^2(2\pi\frac{   k  }L)+ n g^2}\Biggr]. 
\end{multline}
Finally, following eqns. (\ref{eq:ChemicalPlusCommutator}) and (\ref{eq:ChemicalMinusCommutator}), the momentum modes $k^\prime$ ($k$) which minimize (maximize) the chemical potentials need to be found. In the present case ($t_1<0$), these are $k^\prime=0$ and $k=\frac L2$. Thus the resulting chemical potentials are
\begin{multline}
\mu^+_n-\omega= - 2 t + t \delta _{n0}-\sqrt{ t^2 +(n+1) g^2}\\
   +(1-\delta   _{n0}) \sqrt{   t^2+n g^2},
\end{multline}
for any $n$ and 
\begin{multline}
\mu^-_n-\omega=   2 t -t\delta _{n1} - \sqrt{t^2+n g^2}\\
    +(1-\delta   _{n1}) \sqrt{t^2+ (n-1) g^2},
\end{multline}
for $n>0$. A closed form for the critical hopping can be found, but is rather lengthy and will therefore be skipped.\\

We now compare our analytic results to various numerical calculations. Figure \ref{fig:DMRG} shows both analytic approximations along with numerical data from DMRG  \cite{Rossini2007} and mean field \cite{Greentree2006} calculations, where the modified hopping amplitude in the mean field Hamiltonian (\ref{eq:JCHMeanField})  evaluates as $\widetilde J = t$. From the figure,  it can be seen that the effective model gives a much better agreement with the numerical DMRG data, especially the slopes of the lobes agree perfectly at small hopping. The fermion approximation overestimates the size of the Mott lobe. In particular while the lower boundaries are rather well reproduced the upper boundaries have the
wrong slope. Surprisingly though the critical hopping amplitudes seem to agree better with the DMRG data than the
results obtained from the effective strong coupling Hamiltonians. Although the fermion approximation is quantitatively worse than the effective strong-coupling Hamiltonians, it provides a simple approximative solution to the JCHM 
beyond the mean field level which has the advantage of giving a closed form of the ground state.

\subsection{Linear ion chain}

As a second  example  we consider a linear string of 
ions in an ion trap \cite{Ivanov2009}, where the ions
are coupled to an external laser field and 
interact with each other due to Coulomb repulsion via phonon exchange. This system is well described by 
a modified JCH model with a specific
short range hopping with negative effective mass and site dependent parameters.  First we will shortly introduce the model and give a derivation of the corresponding homogeneous limit. Afterwards we will apply the both approximations and discuss the phase boundaries within these approximation giving explicit analytic results for them.\\

As shown in \cite{Ivanov2009}, the Hamiltonian of a linear string of $L$ ions simultaneously
irradiated by a laser which is tuned close to the red radial motional sideband and in the
Lamb-Dicke regime is given by
\begin{multline}
  \hat H =  \sum_{j=0}^{L-1} \omega_j\hat a_j^\dagger \hat a_j + \Delta \sum_j \hat\sigma^\dagger_j 
\hat\sigma^-_j + g\sum_j
\left(\hat \sigma ^\dagger_j \hat a_j + \hat a_j^\dagger \hat\sigma_j^-\right) \\
 + \sum_{j} \sum_{d=1}^{L-j} t_{j,j+d} \left( \hat a^\dagger_{j+d} \hat a_{j} + \hat a^\dagger_{j} \hat 
a_{j+d}\right).\label{eq:JCHIonChain}
\end{multline}
Here $\hat a_j^\dagger$ and $\hat a_j$ describe the creation and annihilation of a local phonon at the $j$th site (ion),
$\hat \sigma_j^\pm$ are the spin flip operators between the internal states of the ion, $\Delta$ is the detuning
of the external laser field from the red motional sideband. $g$ describes the phonon-ion coupling in the Lamb Dicke
limit; for a precise definition see \cite{Ivanov2009}. 
The local oscillation frequencies $\omega_j$ and the hopping amplitudes $t_{j,j+d}$ are determined by
the longitudinal and transversal trap frequencies $\omega_z$ and $\omega_x$ via
\begin{equation}
 \omega_j = -\frac{\omega_z^2}{2\omega_x} \sum_{\begin{matrix} \scriptstyle l=0\\\scriptstyle l\not = j \end{matrix}}^{L-1} \frac{1}{| u_j - u_l|^3}
 \hspace{0.5cm} 
t_{j, j+d} = \frac{\omega_z^2}{2\omega_x}  \frac{1}{| u_j - u_{j+d}|^3}, \label{eq:Constants}
\end{equation}
where $u_j$ are the equilibrium positions  of the ions \cite{James1998}.
For sufficiently large $L$, the equilibrium positions of the ions at the center are 
approximately equidistant, giving  $u_j = j\, \widetilde u$, with $\widetilde u$ being the distance of two adjacent ions. 

Let us now discuss the limit of a homogeneous chain neglecting any boundary effect. In this limit
eqs. (\ref{eq:Constants}) can be rewritten for $L\to\infty$, yielding position independent phonon energies $\omega_j\equiv -\omega$ and hopping amplitudes $t_{j,j+d}\equiv t_d$
\begin{alignat}{3}
t_d &= \frac{\omega_z^2}{2\omega_x \widetilde u^3} \frac 1 {d^3}&&=t\ \frac 1 {d^3} \label{eq:hopping},\\
 \omega&= 2\frac{\omega_z^2}{2\omega_x \widetilde u^3} \zeta(3)&&=2 t \  \zeta(3) \label{eq:omega},
\end{alignat}
where $t = \frac{\omega_z^2}{2\omega_x \widetilde u^3}$ acts as a small parameter and $\omega>0$. $\zeta(x)$ is the Riemann $\zeta$-function. 

\begin{figure*}
\epsfig{file=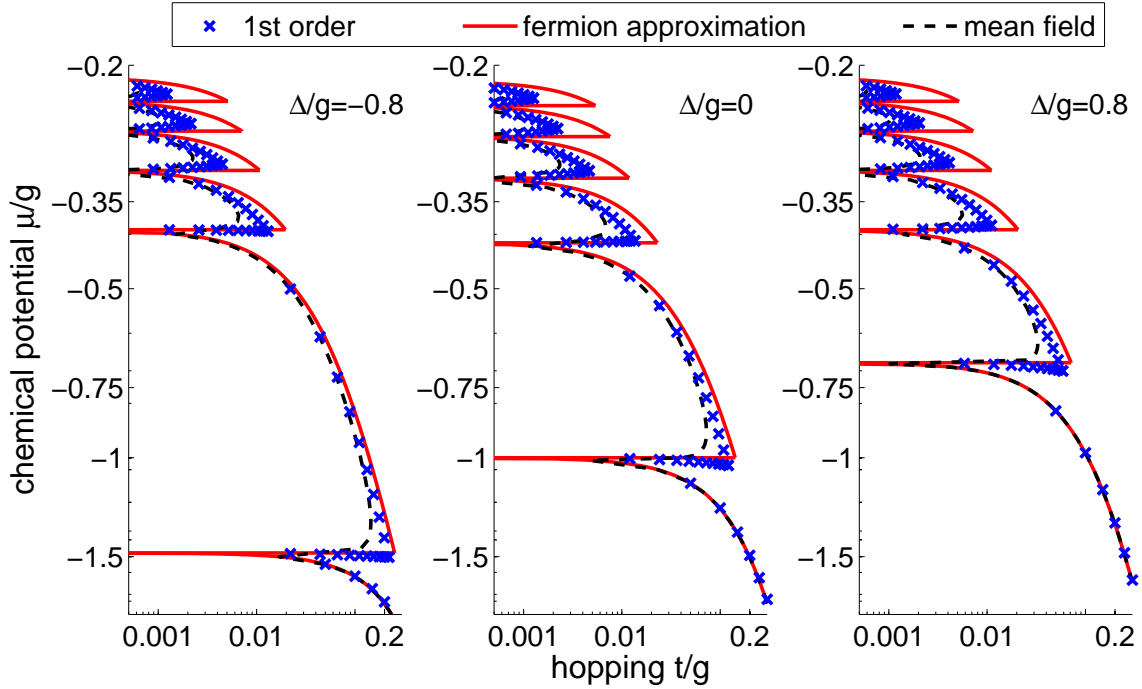,width=2 \columnwidth}
 \caption{(Color Online) Phase diagram of the JCH model for a linear ion chain for three depicted values $\Delta/g=-0.8,0,0.8$. Shown are the upper boundary of the zero filling lobe (always lowest line)  and the boundaries of the lobes with filling from 1 to 5 on a double logarithmic scale. Beside the used approximations (solid line: fermion approximation, crosses: 1st order effective theory) the results from the mean field theory (dot-dashed line) after the canonical transformation are shown. It can be seen, that the fermionic approximation again overestimates the phase boundary (compared to the more reliable effective strong coupling theory) but gives a better agreement compared to the mean field theory (mind the logarithmic scale).
 }\label{fig:PhaseDiagram}
\end{figure*}

One notices a negative oscillator energy $-\omega$ and a negative effective mass, which is a result
of the positive hopping strength $t$.  This negative mass is the reason, why the application of the mean field theory is not that straight forward. When simply calculating the modified hopping amplitude $\widetilde J=-t\sum_d  \frac1{d^3} = - t \zeta(3)$, the hopping gets negative and therefore mean field theory inapplicable. This problem can be overcome by first applying a canonical transformation to all used operators. The transformation
\begin{equation}
 \hat a_j \mapsto (-1)^j \hat a_j
\end{equation}
for the annihilation operator and accordingly to all the other operators $\hat a^\dagger_j$,$\hat\sigma_j^\pm$, maps the JCH model (\ref{eq:JCHreal}) again to the JCH model, but with $t_d \mapsto (-1)^d t_d$. After this transformation, the modified hopping evaluates to $\widetilde J = - t\sum_d \frac{(-1)^d}{d^3}=3t \zeta (3)/4$ being positive. Now the application of the mean field theory is straight forward, following the usual route.

After having introduced the homogeneous limit of the model, the approximations introduced in section \ref{sec:Approximations}  will both be applied. Starting with the effective strong-coupling theory from section \ref{sec:effective}, the chemical potentials for the upper and lower boundary of the lobes are given by eqns. (\ref{eq:chemicalPotentialfirstPlus}) and (\ref{eq:chemicalPotentialfirstMinus}). The proper momentum modes $k^\prime,k^{\prime\prime}$, which minimize (maximize) the chemical potentials are both found to be $k=L/2$. This results from the negative mass. Due to the complexity of the problem, especially the analytic form of $B_n^-$ and $D_n^-$, analytic representation of the chemical potentials are left out here. They can be found straight forwardly just as in the case of the simple cubic JCH model.
\\

\begin{figure}
\epsfig{file=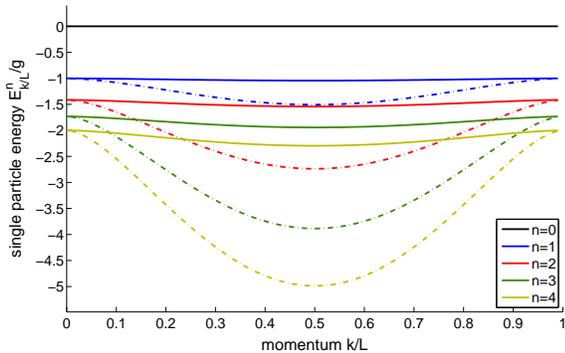,width=\columnwidth}
 \caption{(Color Online) Energies of the JCH Hamiltonian 
for fixed filling $n$ as function of momentum $k$. Shown are the energies from eq. (\ref{eq:JCEnergy}) for the five lowest fillings $0\dots4$ (from top to bottom) for $\Delta=0$ and $g=1$. Solid lines: $t/g=0.02$; Dashed lines: $t/g= 0.2$. One clearly recognizes the minimum at $k=L/2$ and the flat dispersion for $t/g\to0$.}\label{fig:EnergiesFilling}
\end{figure}

When following the approximative method from section \ref{sec:commapprox},  the Hamiltonian for the uncoupled JC models is given by eq. (\ref{eq:JCHMomentum}),  with the phonon energies being
\begin{gather}
\omega_k = -\omega + 2 t \sum_{   d} \frac{ \cos(2\pi\frac{   k    d}L)}{d^3},
\end{gather}
according to (\ref{eq:PhononEnergies}).
Note that since $\omega = 2 t\, \zeta (3)$,  all $\omega_k$'s are negative.
Using the polylogarithm $\text{Li}_n\left(x\right)=\sum_{d=1}^\infty \frac{x^d}{d^n}$ one can write 
them in the explicit form
\begin{eqnarray}
\omega_k= t\left[ \text{Li}_3\left(e^{2 \pi i \frac kL }\right)+\text{Li}_3\left(e^{-2 \pi i \frac kL }\right)-2\zeta(3)\right]. \label{eq:omegak}
\end{eqnarray}
The minimum value of $\omega_k=-7 t \zeta(3)/2$ is attained for $k=\frac{L}{2}$ as expected from the positive sign of the hopping term. The energies for each momentum mode are given by the solution  (\ref{eq:JCEnergyMomentum})  of the JC model and the corresponding spectrum is shown in fig. \ref{fig:EnergiesFilling}.\\

\begin{figure}
\epsfig{file=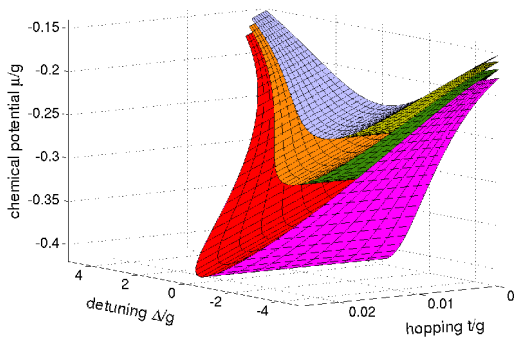,width=1\columnwidth}
 \caption{(Color Online) Phase diagram of the JCH model for a linear ion chain from the fermion approximation. Boundaries of the Mott-insulating lobes (from bottom to top) for $n=2,3,4$. The lobes $n=0$ and $n=1$ are not displayed since they are unbound for $\Delta \to -\infty$.}\label{fig:FullPhaseDiagram}
\end{figure}

From the knowledge of the dispersion relation for different fillings, it is now easy to construct the phase diagram. As discussed in section \ref{sec:commapprox}, the flat dispersion for $t=0$ leads to the ground state having an equal number of excitations in every momentum mode $k$. The chemical potentials for $t>0$ are then determined by the $k^\prime$ and $k$ values, minimizing or maximizing  eqns. (\ref{eq:ChemicalPlusCommutator}) and (\ref{eq:ChemicalMinusCommutator}). When looking at the dispersion in figure \ref{fig:EnergiesFilling}, one recognizes
that this is given for $k^\prime=L/2$ and $k=0$. So, the chemical potentials are given by
\begin{align}
  \mu_n^+ &=  E_{\frac L2}^{n+1}-E_{\frac L2}^{n} \\
  \mu_n^- &=  E_{0}^{n}-E_{0}^{n-1},
\end{align}
and when using the analytic form, eqs. (\ref{eq:JCEnergyMomentum}) and (\ref{eq:omegak}), the 
phase boundaries of the $n$th Mott lobe read
 \begin{multline}
  \mu^+_n = \frac{1}{2} \Biggl[ - \sqrt{4 (n+1)   g^2+\left(\frac{7}{2} \zeta (3) t +\Delta   \right)^2}\\
  -\frac{7}{
1+\delta_{n0}} \zeta (3) t +\delta_{n0}\Delta \\
  +(1-\delta_{n0})\sqrt{4 n g^2+(\frac72 \zeta (3) t + \Delta   )^2}\Biggr],
  \end{multline}
   \begin{multline}
   \mu^-_{n} = \frac{1-\delta_{n1}}{2} \sqrt{4 (n-1) g^2+\Delta ^2}\\
   -\frac{1}{2} \sqrt{4 n   g^2+\Delta ^2} +\frac{\delta_{n1}}{2}\Delta.
\end{multline}

Figure \ref{fig:PhaseDiagram} shows the resulting phase diagram for three values of $\Delta$ comparing the different approaches.  One recognizes the typical lobe structure of the MI-phases with a closing of the lobes at some value $t^{\rm crit}_n(\Delta)$. Whilst the mean field results underestimates the extent of the MI regions, our fermionic approach overestimates them but with a better agreement with the first-order effective strong coupling model compared to the mean field solution.  The main advantage of the fermionic approximation is the easy closed form for the chemical potentials as well as the for the ground state and a more reasonable agreement of the critical hopping amplitude $t^{\rm crit}_n(\Delta)$ as can be seen from the figure. Figure \ref{fig:FullPhaseDiagram} shows the full phase diagram of the model as a function of the detuning $\Delta$ obtained from the fermionic approximation only.\\

\begin{figure}
\epsfig{file=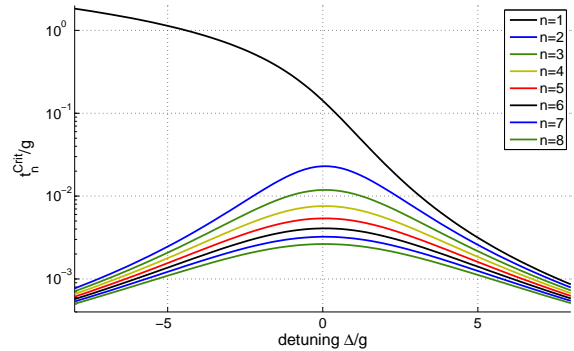,width=1\columnwidth}
 \caption{(Color Online) Critical hopping amplitude $t_n^{\rm Crit}(\Delta)$ giving the point where the MI to 
superfluid transition takes place. From top to bottom: $n=1\dots8$, all for g=0.05}\label{fig:JCrit}
\end{figure}

The critical hopping amplitude $t^{\rm crit}_n(\Delta)$ can easily be calculated from the analytic expressions for the chemical potential
given above. Figure \ref{fig:JCrit} shows the dependence of the critical hopping amplitude from the detuning $\Delta$
for the different MI lobes. One recognizes the unboundness of the first lobe, i.e. $t^{\rm crit}_n(\Delta)\to\infty$ as $\Delta\to-\infty$.\\

\section{Summary}

In summary we have presented two simple analytic approximations to the phase diagram of the Jaynes-Cummings-Hubbard. The first approximation describes the particle-hole excitations in the vicinity of the Mott-insulator to superfluid
transition for a specific filling by a simple effective spin model which generalizes the know results to arbitrary short range hopping. The second approximation treats the spins
as fermions which allows for a simple solution of the model by means of a Fourier transformation. A comparison of both methods to DMRG and mean field data shows reasonable agreement to the numerics.
The approximative description by effective strong-coupling Hamiltonians makes very good quantitative predictions for 
the phase boundaries of the Mott-insulating lobes for small hopping and can be straight forwardly written down up to 2nd order. The fermion approximation performs
also very well for the lower boundaries but is less accurate for the upper ones. It does make
however rather good predictions for the critical hopping at commensurate fillings and has the advantage of giving a closed form for the ground state in the whole parameter regime.  Altogether both methods provide quite reasonable results for the phase boundaries compared to numerical results from DMRG simulations.

\section*{Acknowledgments}

This work has been supported by the TMR network EMALI of the European Union and the
DFG through the SFB-TR 49, the Bulgarian NSF grants VU-F-205/06, VU-I-301/07, D002-90/08 and the excellence program of the Landesstiftung Baden-W\"urtemberg. The authors thank E. Irish and M. Hartmann for pointing out an error in a previous version
of this manuscript and D. Rossini and R. Fazio for providing their DMRG results shown in fig. \ref{fig:DMRG}.

 \bibliographystyle{apsrev}
 \bibliography{BibTex_Alexander_Mering}

\end{document}